\documentclass[epj,twocolumn]{webofc}
\usepackage[varg]{txfonts}   
\usepackage{graphicx}

\wocname{epj}
\woctitle{Seismology of the Sun and the Distant Stars 2016}
\begin{document}
\def\Kepler{\textit{Kepler}\xspace}
\def\Hermes{\textsc{Hermes}\xspace}
\def\gpsc{$\gamma$\,Psc\xspace}
\def\ttau{$\theta^1$\,Tau\xspace}
\def\asterix{KIC\,9163796\xspace}

\newcommand{\Figure}[1]{Figure\,\ref{#1}\xspace}
\newcommand{\Table}[1]{Table\,\ref{#1}\xspace}
\newcommand{\Equation}[1]{Equation\,\ref{#1}\xspace}
\newcommand{\Section}[1]{Section\,\ref{#1}\xspace}

\newcommand{\new}[1]{\textbf{#1}\xspace}
\title{Constraining stellar physics from red-giant stars in binaries $-$ \\ stellar rotation, mixing processes and stellar activity}
\author{P.\,G. Beck\inst{1}
\and T. Kallinger\inst{2}
\and K.~Pavlovski\inst{3}
\and A. Palacios\inst{4}
\and A.~Tkachenko\inst{5}
\and R.\,A.~Garc{\'i}a\inst{1}
\and S.~Mathis\inst{1}\\
E.~Corsaro\inst{1}
\and C.~Johnston\inst{5}
\and B. Mosser\inst{6}
\and T.~Ceillier\inst{1} 
\and {J.-D.~do\,Nascimento Jr.\inst{7,8}} 
\and G.~Raskin\inst{5}
}
\institute{Laboratoire AIM Paris-Saclay, CEA/DRF --- CNRS --- Universit\'e Paris Diderot, \\IRFU/SAp Centre de Saclay, F-91191 Gif-sur-Yvette Cedex, France 
\and Institut f\"ur Astronomie der Universit\"at Wien, T\"urkenschanzstr. 17, 1180 Wien, Austria 
\and Department of Physics, Faculty of Science, University of Zagreb, Croatia 
\and LUPM, Universit\'e Montpellier II, Place Eug\'ene Bataillon cc -0072, F-34095, Montpellier cedex 5, France
\and Instituut voor Sterrenkunde, KU Leuven, 3001 Leuven, Belgium 
\and LESIA, CNRS, Universit\'e Pierre et Marie Curie, Universit\'e Denis Diderot, Observatoire de Paris, 92195 Meudon cedex, France
\and Harvard-Smithsonian Center for Astrophysics, Cambridge, MA 02138, USA
\and Departamento de F\'isica Te\'orica e Experimental, Universidade Federal do Rio
}
\abstract{
The unparalleled photometric data obtained by NASA's \Kepler Space Telescope has led to an improved understanding of stellar structure and evolution - in particular for solar-like oscillators in this context. Binary stars are fascinating objects. Because they were formed together, binary systems provide a set of two stars with very well constrained parameters. 
 Those can be used to study properties and physical processes, such as the stellar rotation, dynamics and rotational mixing of elements and allows us to learn from the differences we find between the two components.  In this work, we discussed a detailed study of the binary system \asterix, discovered through \Kepler photometry. The ground-based follow-up spectroscopy showed that this system is a double-lined spectroscopic binary, with a mass ratio close to unity. However, the fundamental parameters of the  components of this system as well as their lithium abundances differ substantially. \Kepler photometry of this system allows to perform a detailed seismic analysis as well as to derive the orbital period and the surface rotation rate of the primary component of the system. Indications of the seismic signature of the secondary are found.  The differing parameters are best explained with both components located in the early and the late phase of the first dredge up at the bottom of the red-giant branch. Observed lithium abundances in both components are in good agreement with prediction of stellar models including rotational mixing.
 By combining observations and theory, a comprehensive picture of the system can be drawn.
} 
\maketitle
\section{Introduction}
\vspace{-1mm}

\begin{figure*}[t!]
\centering
\resizebox{0.9\textwidth}{!}{  \includegraphics{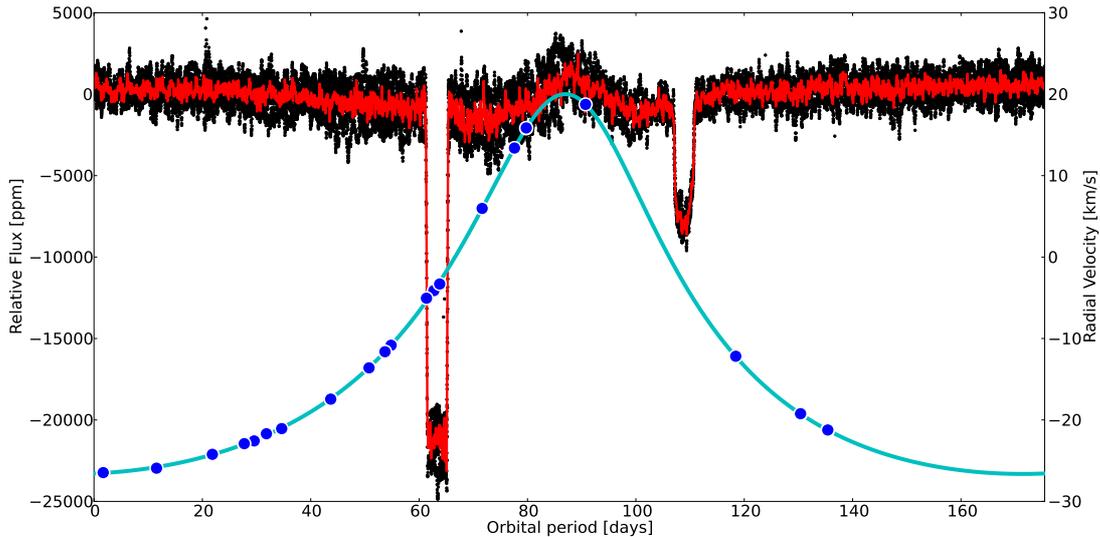} }
\caption{ Combined phase diagram for the eclipsing system KIC\,9540226 observed using
space photometry from the \Kepler satellite and radial velocities from ground-based
spectroscopic observations with the Hermes spectrograph. The \Kepler
photometry is presented on the left vertical axis. Black data points are the
original measurements, while the red line shows the averaged light curve, which
was determined by rebinning the phase diagrams into 30 minutes bins. The
deviation from the averaged light curve is caused by solar-like oscillations. The
radial velocities are shown as solid, blue dots (right vertical axis). The solid
blue line represents the orbital model of a Keplerian orbit. (Figure from \cite{BeckPhD})\label{fig:phaseDiagram}}
\end{figure*}

Asteroseismology is a powerful tool to study stellar structure and dynamics as well as their evolution \citep{Aerts2010}. The NASA \Kepler mission \citep{Borucki} delivered unprecedented, high-precision photometric data, enabling us to perform detailed asteroseismic analysis for thousands of stars \citep[e.g.][]{Kallinger2010,Hekker2011,Chaplin2011,Chaplin2014,Huber2016}. 
Seismology provides important inferences on the internal structure of stars such as the evolutionary state \citep[e.g.][]{Bedding2011,Mosser2015, Vrard2015,Kallinger2016}, stellar rotation \citep[e.g.][]{Beck2012,Mosser2012,Deheuvels2012,Deheuvels2014,Deheuvels2015,diMauro2016}, or the characterisation of members of open clusters  \citep{Miglio2012,Corsaro2012,Brogaard2016,Epstein2014}. Also distances  can be estimated through asteroseismology \citep[e.g.][]{Mathur2016} and now can be compared to distances from trigonometric parallaxes, provided by GAIA \citep[e.g.][]{deRidder2016}. Such wealth of information allows us to put strong constraints on the interpretation of observations of stars and view the star in a wider context.  Asteroseismology unfolds its full potential of precision when combined with fundamental parameter from spectroscopy. This allows an accurate determination of the mass and radius of the star \citep{Chaplin2014,Lebreton2014}. 
A particular treasure trove for stellar astrophysics are oscillating stars in binary systems. For such stars we have the unique possibility to constrain radius and mass 
with two different independent methods \citep[e.g.][]{TorresReview,Frandsen2013}. Indeed, comparing masses from binary solutions and stars in cluster, evidence was found that the seismic scaling relations \cite[e.g.][]{Chaplin2014} are systematically overestimating the stellar mass by about 15\% \citep{Gaulme2016,Epstein2014}. 

Binary systems are rewarding to study, since many parameters are well constrained (e.g. mass, radius, metallicity). Unless a binary system was created through a rare event of capturing, both  components were born together, they have the same age and metallicity.  For stars in binary systems,  the only differences between the two components are the difference in mass and eventually from the surface rotation if the system is synchronised.  The difference in mass then translates into different speeds of stellar evolution \cite[e.g.][]{Kippenhahn}. In case of a double-lined spectroscopic binary system (SP2), this parameters can precisely be determined from the ratio of the radial velocity amplitudes, even for unknown inclinations \citep{TorresReview}.  As we will demonstrate later in this proceedings paper, the SB2 feature, actually allows to measure mass ratio between the two components with an accuracy of about 1\%. Using SB2 thus allows a model independent approach. If one or both components show stellar oscillations, seismic techniques can be applied to further study the stellar structure of the oscillating component in detail. With such a well constrained set of parameters, one can draw a comprehensive picture of the full system and its components.

By now about 3000 stars, i.e. $\sim$1.3\% of all stars observed by \Kepler, are found to be binaries \cite[and references therein]{Kirk2016}. About 40 systems with an oscillating red-giant component are known and characterised  whereby some studies focused on the eclipsing systems \cite{Gaulme2013, Gaulme2014}, while another presented a catalogue of red giants in binary systems with tidally induced flux modulation \cite{Beck2014a}. A set of additional candidates of red giants in binaries was proposed by \cite{BeckToulouse}. Finally, binaries with solar-like oscillators on the main-sequence were found \citep[e.g.][]{Appourchaux2015,White2016,Beck2016b}.

Several binaries with a red-giant component have been analysed in detail using \Kepler photometry and ground-based spectroscopy: KIC\,8410637 \citep{Hekker2010,Frandsen2013,Themessl2016}, KIC\,5006817 \citep{Beck2014a}, KIC\,9246715 \citep{Rawls2016}, KIC\,5640750 \citep{Themessl2016}, and KIC\,9540226 \citep{Themessl2016}.
With this work, we added KIC\,9163796 \citep{Beck2016} (also nicknamed \textit{Asterix\,\&\,Obelix})  to the sample of well studied binary systems. 

In this paper, we first discuss in Section\,\ref{sec:observations} the different types of observational features that can be used to detect binaries and highlight the benefit of ground-based spectroscopy. The \Kepler photometry and \textsc{Hermes} spectroscopy of the system of \asterix is described in detail in Section\,\ref{sec:Asterix}. The seismology of the primary and secondary component of \asterix are discussed in Section\,\ref{sec:seismology}. The indicators of stellar activity are described in Section\,\ref{sec:activity} and the measured and predicted abundances of lithium are compared in Section\,\ref{sec:modelling}. The findings are summarised in Section\,\ref{sec:conclusions}. 

\section{Observations of Binaries \label{sec:observations}}
\subsection{\Kepler photometry}
Given the unprecedented quality of the photometric data provided by the \Kepler space mission, binaries can be detected by searching for four different features. First, there are the classical eclipsing binary systems, which can be detected through the periodic eclipses. A second class of binaries are the ellipsoidal variables, in which the flux modulation originates from tidal interaction \cite[e.g.][]{Welsh2011,Fuller2012,Thompson2012}. A special case of these ellipsoidal variables are the eccentric cases, in which the flux modulation only occurs during the periastron passage. This leads to a periodic modulation of the observed object brightness. The characteristic shape of the light curve at periastron triggered the colloquial but controversial nick name of \textit{'heartbeat stars'}. This term, coined by \citep{Thompson2012}, was debated during this conference.  
Physically, eccentric eclipsing binaries and heartbeat stars are the same and thanks to the photometric quality of \Kepler even combinations of eclipses and heartbeat events are found \citep[e.g. KIC\,2697935, KIC\,9540226, KIC\,10614012; in][see also Figure\,\ref{fig:phaseDiagram}]{Beck2014a,BeckPhD}. 
The only difference between the two sets is that eclipses allow a better determination of the inclination, though are restricted to a very narrow degree ranges, while ellipsoidal variables can be found at nearly all inclination. Therefore, separating heartbeat stars from eclipsing binaries is unpractical and unnecessarily narrowing down sample sizes. 
The third and fourth way of detecting binaries are not found in the light curves, but in the Fourier space. On one hand, it was shown that oscillating stars in binaries can be identified from the phase modulation of coherent modes \cite{Murphy,Kurtz}. However, the stochastic nature of solar-like oscillations and the granulation background make this approach impracticable for solar-like and red-giant stars. Still binaries can be detected from the oscillation spectrum, if both components are visible. These seismic binaries (hereafter in analogy to the double-lined spectroscopic binaries, SB2, also referred to as PB2), however can only be found in a very narrow mass range and have smaller amplitudes due to photometric dilution \citep{Miglio,Johnston2017}.

On the other hand, an additional phenomenon that in principle is present in binaries is the so-called \textit{Doppler beaming} \cite{LoebGaudi2003,Zucker2007}. This effect originates from the varying radial velocity in the binary, periodically shifting the stars' spectral energy distribution with respect to \textit{Kepler's} photometric pass band and an alternated arrival rate of photons. In a single-lined binary (SB1), a net increase or decrease would be observed if the primary component moves toward or away from the observer, respectively. 
While this effect was predicted in 2003 \cite{LoebGaudi2003,Zucker2007},  
its existence could only be shown from high-quality photometry  provided by the \textsc{CoRoT} and \Kepler space missions for binary systems with large radial-velocity amplitudes \citep[e.g.][]{Bloemen2011}. Due to the combination of the red-biased spectral energy distribution of red-giants and the pass band of \Kepler, Doppler beaming should also be observable for binaries with red-giant components. In particular, for the low-luminosity red-giant star KIC\,5006817, models suggest that  the net amplitude of this effect over the full orbital cycle should be about 300\,ppm~\cite{Beck2014a}. Therefore, this effect is influencing the light curve and needs to be taken into account for its modelling. However, finding a small effect in binaries with periods compatible or longer than the length of a \Kepler data quarter (90\,days) is challenging, as it is nearly impossible to distinguish it from instrumental noise and therefore will be filtered out in the data calibration.

\subsection{Ground-based spectroscopy}

In addition to the photometric or seismic detection of binaries, the analysis of multiple star systems requires the combination of space photometry with high-resolution time-resolved spectroscopy. The most fundamental task of ground-based follow-up observations is to obtain an independent confirmation of the binary system itself. 
{This is particularly important for systems where the contamination through neighbouring field stars or some combination of activity or instrumental systematic effects could mimic the signature of stellar binarity. }
Only periodic radial velocity (RV) variations can confirm such systems.
Having all phases of the orbit well covered with observations further allows to derive the orbital elements. Although it is in principle possible to derive all orbital parameters from the average light curve, it is beneficial, especially in case of non-eclipsing heartbeat stars, to narrow down the used parameters space by including the orbital parameters from spectroscopy. This is particularly true for ellipsoidal variables. 

\begin{figure}[t!]
\centering
\resizebox{0.95\columnwidth}{!}{  \includegraphics{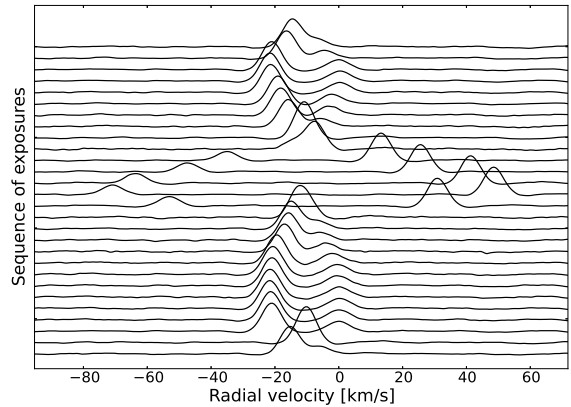} }
\caption{ Time series of cross-correlation profiles of individual spectra of \asterix with a red-giant-star template. The higher and the smaller peak corresponds to the average line profile of the primary and secondary component in the system.
\label{fig:SB2curve}}
\end{figure}

\begin{figure*}[t!]
\centering
\resizebox{\textwidth}{!}{  \includegraphics{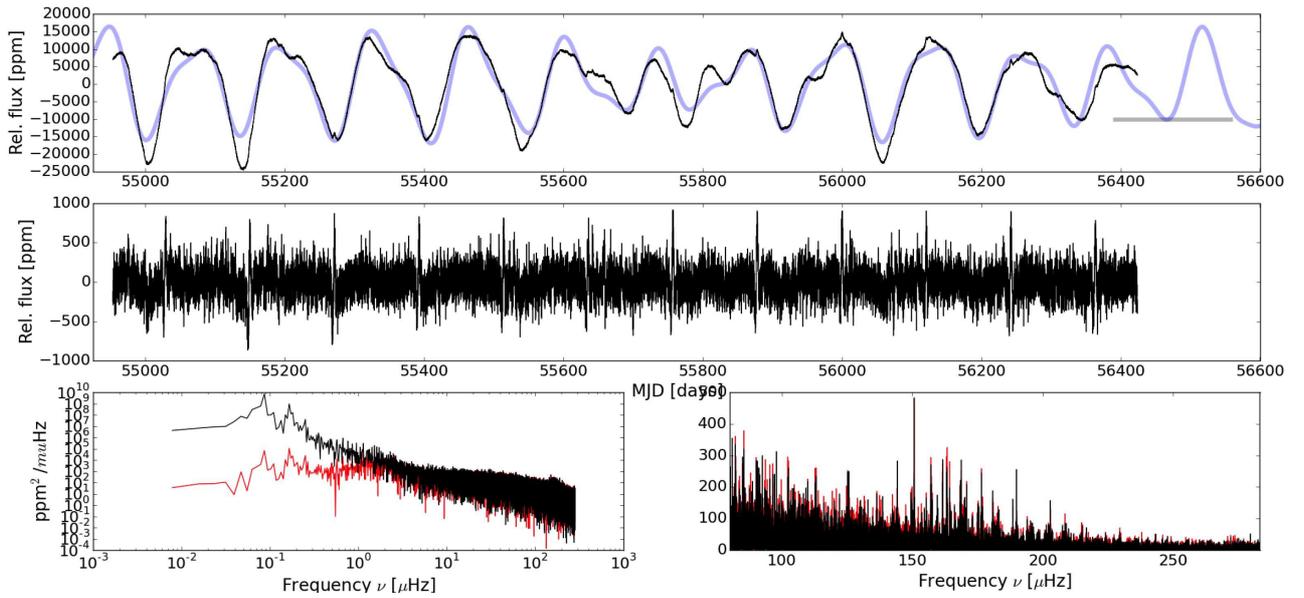} }
\caption{ Light curve of \asterix. The top and lower panel show the full light curve (Q0-Q17), smoothed with a filter of 150 and 4 days, respectively. The synthetic light curve, depicted as solid light grey line is  from the four significant frequencies found through a prewhitening procedure.  The time base of the spectroscopic monitoring is indicated through the horizontal grey bar. The bottom left and right panel show the power spectral density in logarithmic and linear scalings, whereby the red and black curves represent the spectra originating from the inpainted \Kepler data set with a smoothing of 4 and 150\,days, respectively.
\label{fig:LCcurve}}
\end{figure*}

In the optimal case, a binary system is found to be a double-line binary. In such a system the stellar spectra of both components are moving in anti-phase to each other. The RV amplitudes of such systems provide directly the mass ratio. 
In addition, having the information on the inclination allows us to obtain the dynamical masses of the system \cite{TorresReview}.

Furthermore, using all the spectra obtained to measure the RV of the system, allows to substantially increase the signal-to-noise (S/N) in the combined spectrum. Such spectrum can be used to perform a detailed spectroscopic analysis of one or both components of the system. Especially for SB2s', several methods have been developed to search and extract the individual stellar spectra from the composite spectrum, \cite[e.g. \textsc{TodCor} or \textsc{FDBinary};][ respectively]{TodCor,Ilijic2004}.  Recent advances in the analysis techniques and data quality have  pushed the limit for a positive detection below 1 percent of contribution of the secondary or even tertiary component to the total flux of the system \cite[e.g.][]{Kolbas2015,Beck2017Theta}. Even the non-detection of the secondary is a valuable information, as it sets the upper limit for luminosity of the \hbox{secondary \citep[e.g.][]{Beck2014a}.}

\section{ The red-giant/red-giant SP2  system KIC\,9163796 \label{sec:Asterix}}

In a previous study \cite{Beck2014a}, it was shown that KIC\,9163796 is  a non-eclipsing binary system, though it exhibits clear ellipsoidal flux modulation every 121.3\,days during periastron (see Figure\,\ref{fig:LCcurve}). The analysis of radial velocities from ground-based spectroscopy of this system confirmed that \asterix is an eccentric binary system ($e$$\simeq$0.7). The primary component  shows an oscillation power excess at about 160\,$\mu$Hz. 
A visual inspection of the cross-correlation profiles of the time series revealed that this system is an impressive case of an SB2 system \cite{BeckPhD}. The time series of the individual responses of the cross correlation of the stellar spectra with a red-giant template is depicted in Figure\,\ref{fig:SB2curve}.
In this section, we are summarising the results of our study, dedicated to obtain a comprehensive and complete picture of \asterix~\cite{Beck2016b}.

\subsection{Long period flux modulation}

\begin{figure*}[t!]
\centering
\resizebox{0.60\textwidth}{!}{  \includegraphics{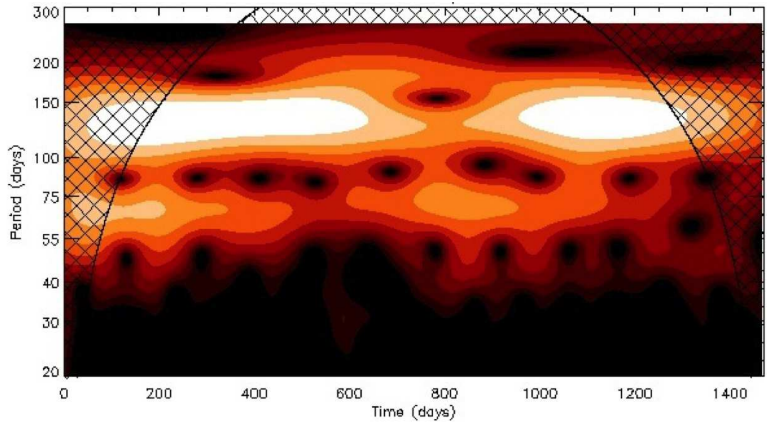} }
\caption{\label{fig:surfaceRotation}
Analysis of the long periodic light curve modulation of KIC9163796 in the period range between 20 and 200 days. The left panel depicts the wavelet analysis of the full time series in period space. Period regions in the wavelet for which the window function is larger than the covered time base are indicated. The right panel shows the average signal over the full time base. }
\end{figure*}

A careful rereduction of the full available \Kepler data set (Q0-Q17), as described in  \cite{Garcia2011,Garcia2014}, reveals that this system exhibits strong flux modulations (see top panel of Figure\,\ref{fig:LCcurve}). Among the red-giant stars observed with \Kepler, photometric rotational variability is found only for a small fraction of stars. In a very recent study, only $\sim$300 red-giant stars were found to exhibit surface modulation in the range of a few to several hundred days \citep{Ceillier2016RedGiants}.  Applying optimised techniques \cite{Garcia2014Rotation,Ceillier2016KOI} to the light curve of this system, such as the wavelet analysis depicted in \Figure{fig:surfaceRotation}, we find the period of the modulation to be $\sim$130\,d. 

The photometric period, likely originating from stellar rotation of the primary component  is close to the orbital period of $\sim$120\,d. By comparing with values of solar and anti-solar differentially surface rotation reported in the literature \cite[e.g.][]{Kovari2015,Kuestler2015} we cannot reproduce this discrepancy of 10\,d. Therefore, rotation of the primary component and the orbital motion are currently not synchronised. One might speculate that the system was in a synchronised state on the main sequence, but the rapid changes of the stellar structure and radius in particular could have lifted this state or that the period of the system is to long and the system too wide to reach a perfect synchronisation (see \cite{Verbunt1995} for a detailed discussion of the tidal evolution of binaries containing red-giant stars).

\subsection{Spectroscopic study \& fundamental parameters \label{sec:SB2analysis}}
Since the identification of \asterix as a binary, this system was monitored with the \textsc{Hermes} spectrograph \citep{Raskin2011,RaskinPhD}, mounted to the 1.2\,m \textsc{Mercator} telescope on La Palma, Canary islands in 2013 \cite{Beck2014a} and 2015 \cite{Beck2016}. The cross-correlation response function of the individual spectra, depicted in Figure\,\ref{fig:SB2curve}, show that this system is an SB2 with strong contributions of both components. 
For this analysis of the SB2 feature, we applied spectral disentangling in the Fourier domain \citep{Hadrava1995}  with the {\sc FDBinary} code \citep{Ilijic2004} to search for the optimal set of orbital elements. Experience has shown that the best region in the spectrum to derive a consistent orbital solution is located at 520\,nm. This range includes many prominent absorption lines and enough regions to determine the pseudo continuum which is crucial for the normalisation of the spectra.  Because the signature of the secondary is clearly visible in the spectrum, the initial set of parameters is straightforward to define and the orbital period is well defined from the analysis of the \Kepler light curve. For details of the full spectral disentangling we refer the reader to \citep{Beck2016}. 

The RV semi amplitudes are $K_1$\,=\,35.25$\pm$0.12\,km/s and $K_2$\,=\,35.77$\pm$0.13\,km/s for the primary and secondary, respectively. These radial velocity amplitudes translate into a mass ratio, $q$\,=\,$M_1/M_2$\,=\,1.015$\pm$0.005. Therefore, the SB2-nature of this system allows us to measure  mass ratio between the two components much more precisely than what could be achieved through seismology. 
The cross-correlation response functions of the individual spectra (Figure\,\ref{fig:SB2curve}) show that average-absorption line profiles of the primary and secondary components are equal in hight in the composite spectrum. This indicates that the primary contributes substantially more flux to the integrated brightness of the system than the secondary. This indicates that the fundamental parameters of both stars are quite different for these stars with a mass ratio close to unity. 

The final orbital solution was used to extract a 20\,nm wide spectral segment around the H$_\beta$ line as well as the {Mg}\,\textsc{i} triplet which were subsequently used to derive the spectroscopic analysis  with the \textit{Grid Search in Stellar Parameters} (\textsc{gssp}\footnote{The GSSP package is available for download at
https://fys.kuleuven.be/ster/meetings/binary-2015/gssp-software-package.}) software package by \cite{Tkachenko2015}. In such an analysis the contribution of both components to the total flux need to be taken into account and correctly normalised to the individual continuum flux. Although the mass ratio derived from the solution of spectral disentangling indicates that the masses of the two stars is very similar the spectroscopic analysis finds that  both components differ substantially in terms of the effective temperature with a difference of 600\,K. Consequently, the primary is substantially more luminous than the secondary and contributes to $\sim$60\% of the total flux. The spectroscopic analysis shows that both components share the same value of the stellar metallicity, [M/H]\,$\simeq$\,$-$0.37\,dex, which is coherent with the fact that both stars were born in the same cloud. 

The main difference between the two components, besides the effective temperature was found in the abundance of lithium (Li). The orbital solution was used to extract a 5\,nm wide segment around the lithium resonance doublet at 670.78\,nm. While the primary  component shows 1.3\,dex, the secondary exhibits a stronger lithium line, leading to 2.6\,dex.

\begin{figure*}[t!]
\centering
\resizebox{0.85\textwidth}{!}{  \includegraphics{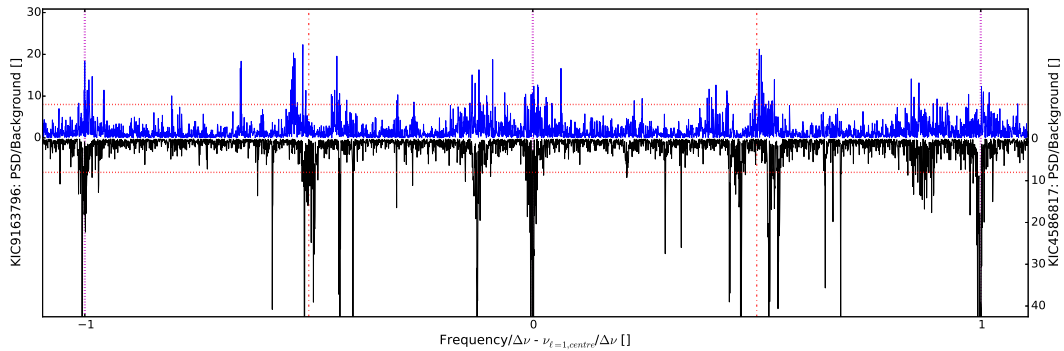} }
\caption{  Normalised power spectral density of \asterix (blue, top spectrum) and KIC\,4586817 (black, bottom spectrum). The background contribution has been removed by division of the PSD through the background model. The frequency was divided by the respective large separation and centred on the central radial mode.
For \asterix, the light grey spectrum depicts the dilution factor corrected spectrum.
The frequencies of the $\ell$\,=\,0 and 1 modes are marked by magenta dotted and red dash dotted vertical lines. The dotted horizontal lines mark the traditional significance threshold of 8 times the background signal.
\label{fig:psdComparison}}
\end{figure*}

\section{Asteroseismic analysis of \asterix \label{sec:seismology}}
From the preceding study of this system it was known that  the oscillation amplitude of the system is very small. The power spectral density (PSD) is shown in the bottom panels of Figure\,\ref{fig:LCcurve}. In the light of the spectroscopic results this is now easy to understand due to the effects of photometric dilution \cite[e.g.][]{Miglio}. Further challenges for the analysis of the PSD are introduced by the regular gaps, originating from on-board manoeuvres of the \Kepler satellite which complicate the spectral-window function \cite{Garcia2011,Garcia2014}. The effect is depicted through the difference of the original and inpainted \Kepler data as red and black curves, respectively in the bottom panels of Figure\,\ref{fig:LCcurve}. These alias frequencies are  dominating the high-frequency regime of the PSD and if not corrected for, the low-amplitude oscillation signal is hidden by the window function. Such alias frequencies are reduced by applying inpainting techniques described in \cite{Garcia2014,Garcia2011}. 

Following the standard approach of fitting several components seen in the PSD of a red giant, (for a summary we refer the reader to \cite{Kallinger2010,BramsPaper}), we obtain the normalised PSD of \asterix shown in the top spectrum of \Figure{fig:psdComparison}. 
From scaling relations \cite[e.g.][]{Kallinger2010,Chaplin2011}, we find an estimate of mass and radius of 1.4\,M$_\odot$ and 5.4\,R$_\odot$, respectively. 

The unambiguous identification of the excited non-radial modes was found to be challenging due to the low S/N in the spectrum. In the case of the binary KIC\,9246715  \cite{Rawls2016}, two overlapping power excesses were found, further complicating the PSD. For that particular binary, both components are found to be located in the secondary clump and therefore are already in the stable phase of helium core-burning. 
The frequency of the power excess and the period spacing of consecutive dipole modes \cite{Beck2011,Bedding2011,Mosser2011} indicate that the more massive primary of \asterix is a low-luminosity red-giant star in the hydrogen-shell burning phase. In this phase, at the bottom of the red-giant branch, changes of the radius and luminosity are happening on much shorter time scales and therefore, we expected the power excesses of the two stars to be well separated. To be sure that all peaks belong only to the primary component and to guide the eye for the mode identification, the PSD of a single red-giant star is shown. This comparison of \asterix to KIC\,4586817, depicted in Figure\,\ref{fig:psdComparison}, shows a very good resemblance and was guiding the eye in the initial mode identification. 

From the mass ratio between the two components and the fundamental parameters of the secondary of this system, it is expected that also the less massive star should be found to oscillate. However, the secondary power excess would be hard to detect due to small amplitudes and the effects of photometric dilution. Based on the Bayesian evidence we find that a model with two Gaussian components significantly better fits the observed PSD than a model with only one Gaussian.  The second power excess is located at $\sim$215\,$\mu$Hz. This value is in qualitative good agreement with the estimated position of the secondary power excess, if scaled from the output parameters of the spectral disentangling analysis.  Unfortunately, the mode visibility of the secondary is too low to determine a reliable estimate for the large frequency separation $\Delta\nu$.

The PSD of \asterix also shows the rotational splitting of dipole mixed modes (see Figure\,\ref{fig:psdComparison}). By using the \textsc{Diamonds} \textit{(high-DImensional And multi-MOdal NesteD Sampling)} code \citep{diamonds}, precise frequency values of these non-radial modes were extracted and the rotational splitting computed. 
Following the standard forward-modelling approach of using a two-zone model of assuming a rigidly rotating core and envelope, we used a representative model, calculated with the \textit{Yale Stelar Evolution Code, YREC} \cite[][]{Demarque2008, Guenther1994}. The rotation rates of both zones was optimised until a good agreement with the set of measured rotational splittings was reached. The seismic result is a surface rotation period of 81\,nHz, which is in good agreement with the main period of the surface brightness modulation of 79\,nHz. In the primary of this system the core rotates about 6.9 times faster than the surface. This is in the lower range of observed rotational gradients between the core and the envelope of 10-30 times \cite{Beck2012,Deheuvels2012,Deheuvels2014,diMauro2016,Goupil2013}, but not exceptionally low. This differs from the quasi-rigid rotation, which was found in red giants, located in the secondary clump \cite{Deheuvels2015}.

\section{Stellar activity \label{sec:activity}}

The light curve, depicted in Figure\,\ref{fig:LCcurve}, indicates that at least one star must show strong stellar activity. The existing observational material allows us to estimate the stellar activity in two different ways. The classical way of measuring activity is the $\mathcal{S}$-index, quantifying the emission line at the core of the Ca\,H\&K lines in the near Ultra-Violet. The spectra of \asterix show a very strong emission, compared to other red giants and solar-analogues, observed with the \textsc{Hermes} spectrograph \cite{Beck2014a,Beck2015,Beck2017Theta,Beck2016Atlas,Salabert2016Analogues}. However, it is very challenging to extract a reliable $\mathcal{S}$-index for a star in a SB2 system with a bright secondary component.

Another proxy is the linewidth of the oscillation modes. Several studies have presented indications that stellar activity also reduces the amplitudes of solar-like oscillations and increases the damping \citep[e.g.][]{Mosser2009,Garcia2010,Bonanno2014,Gaulme2014}. Because the full width at half maximum is independent of photometric dilution through the other stellar component, it can be directly compared to measured mode in other stars. Using the sample of \cite{Corsaro2015} as a benchmark, we see that the full width at half maximum of dipole modes in \asterix is about four times broader than those in normal active stars on average. This can only be explained through strong mode damping.

\begin{figure}[t!]
\centering
\resizebox{0.90\columnwidth}{!}{  \includegraphics{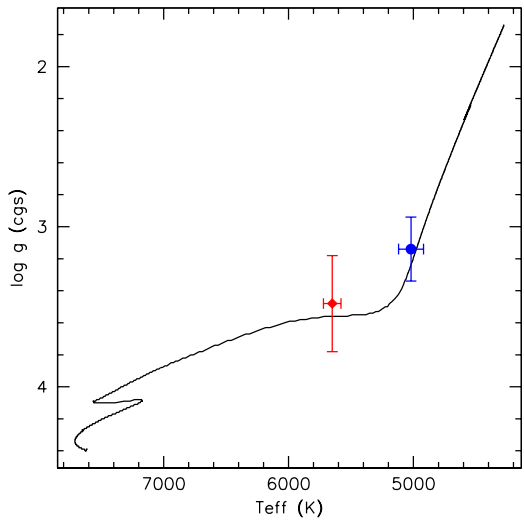} }
\caption{ Evolutionary track for a representative model of both components of \asterix. The current position of the primary and secondary are indicated through the blue and red marker, respectively.
\label{fig:FDU}}
\end{figure}

\section{Stellar modelling \label{sec:modelling}}
 \asterix provides a set of stars, well constrained through photometry,  spectroscopy, and seismology. Therefore, this binary system is a rewarding target for modelling. 
 
 Both components were modelled with the \textsc{starevol} \citep[][]{Decressin2009,Siess2006}. The first result is that the difference in fundamental parameters is in agreement with the position of the primary and secondary being in the late and early phase of the first dredge-up event (FDU) at the bottom of the red-giant branch as illustrated in Figure\,\ref{fig:FDU}. Therefore, both components are framing one of the most important events on internal mixing in stellar evolution. 

In a second step, these models were used to calculate and compare the lithium abundance with various scenarios of rotational histories. The best agreement was found with a (quasi) rigidly rotating progenitor. This is in agreement with findings of modern asteroseismology \citep[e.g.][]{Benomar,Kurtz}

\section{Conclusions \label{sec:conclusions}}
\asterix is an intriguing binary system. Through a multi-technique approach we arrived to the conclusion that both the primary and secondary are oscillating and are located in the late and early phase of first dredge-up event. This is astonishing, because both stars have a very similar mass. Therefore, this system is an impressive example of the how sensitive is the stellar evolution to small differences in mass. An extensive discussion of the results is provided in the main paper \citep{Beck2016}.\\

{\small \textbf{Acknowledgements.} We acknowledge the work of the team behind \textit{Kepler}. Funding for the \textit{Kepler} Mission is provided by NASA's Science Mission Directorate. 
We thank the technical team as well as the observers of the \textsc{Hermes} spectrograph and Mercator Telescope, operated on the island of La Palma by the Flemish Community, at the Spanish Observatorio del Roque de los Muchachos of the Instituto de Astrof{\'i}sica de Canarias. PGB, RAG and BM acknowledge the ANR (Agence Nationale de la Recherche, France) program IDEE (n$^\circ$ ANR-12-BS05-0008) "Interaction Des Etoiles et des Exoplanetes".  PGB, RAG, StM and BM also received funding from the CNES grants at CEA. KP was supported by the Croatian Science Foundation grant 2014-09-8656. E.C. is funded by the European Community's Seventh Framework Programme (FP7/2007-2013) under grant agreement N$^\circ$312844 (SPACEINN). StM acknowledges support by the ERC through ERC SPIRE grant No. 647383. The research leading to these results has received funding from the European Community's Seventh Framework Programme ([FP7/2007-2013]) under grant agreement No. 312844 (SPACEINN) and under grant agreement No. 269194 (IRSES/ASK).}

\end{document}